\documentclass[10pt,journal,compsoc]{IEEEtran}
\usepackage{wrapfig}

\usepackage{graphicx}
\usepackage{amssymb}
\usepackage{subcaption}
\usepackage{multicol}
\usepackage[cmex10]{amsmath}
\usepackage{amsfonts,mathtools}
\usepackage{url}
\usepackage{array}
\usepackage{url}
\usepackage{bm}
\usepackage{graphicx}
\usepackage{mathrsfs}
\usepackage{amsthm} 
\usepackage{bm}
\usepackage{tikz}
\usepackage{cite}
\usepackage{color,soul}
\usetikzlibrary{bayesnet}

\usepackage{adjustbox}
\usepackage{amssymb}
\usepackage{algorithm}
\usepackage{algorithmic}

\usepackage{authblk}
	\title{Variational Bayesian Filtering with Subspace Information for Extreme Spatio-Temporal Matrix Completion}
\author[1]{Charul Paliwal }
\author[1]{Pravesh Biyani}
\author[2]{Ketan Rajawat}
\affil[1]{Indraprastha Institute of Information Technology, Delhi}
\affil[2]{Indian Institute of Technology, Kanpur}

\providecommand{\Diag}[1]{\text{Diag}\left(#1\right)}
\providecommand{\norm}[1]{\left \| #1 \right \|}
\newcommand{\bftab}{\fontseries{b}\selectfont}
\def \a {\mathbf{u}}
\def \p {\bm{p}}
\def \b {\mathbf{v}}
\def \A {\mathbf{U}}
\def \B {\mathbf{V}}
\def \u {\mathbf{u}}
\def \v {\mathbf{v}}
\def \U {\mathbf{U}}
\def \V {\mathbf{V}}

\def \X {\mathbf{X}}
\def \Z {\mathbf{Z}}
\def \x {\mathbf{x}}
\def \Y {\mathbf{X}}
\def \y {\mathbf{x}}
\def \x {\mathbf{x}}
\def \I {\mathbf{I}}
\def \F {\mathbf{F}}

\def \j {\mathbf{f}}
\def \L {\mathbf{L}}
\def \E {\mathbf{E}}
\def \e {\mathbf{e}}

\def \J {\mathbf{F}}
\def \Rn {\mathbb{R}}

\def \Lam {\boldsymbol{\Lambda}}
\def \gam {\boldsymbol{\gamma}}
\def \N {\mathcal{N}}
\def \No {\mathbf{N}}

\def \L {\mathcal{L}}

\def \mub {\boldsymbol{\mu}}
\def \th {\boldsymbol{\theta}}

\def \up {\boldsymbol{\upsilon}}
\def \al {\boldsymbol{\alpha}}
\def \Sig {\boldsymbol{\Sigma}}
\def \Xib {\boldsymbol{\Xi}}

\def \v {\mathbf{v}}
\def \dott {\boldsymbol{\cdot}}

\begin{document}

\maketitle
 
\begin{abstract}

Missing data is a common problem in real-world sensor data collection. The performance of various approaches to impute data degrade rapidly in the extreme scenarios of low data sampling and noisy sampling, a case present in many real-world problems in the field of traffic sensing and environment monitoring, etc. However, jointly exploiting the spatiotemporal and periodic structure, which is generally not captured by classical matrix completion approaches, can improve the imputation performance of sensor data in such real-world conditions.
We present a Bayesian approach towards spatiotemporal matrix completion wherein we estimate the underlying temporarily varying subspace using a Variational Bayesian technique. We jointly couple the low-rank matrix completion with the state space autoregressive framework along with a penalty function on the slowly varying subspace to model the temporal and periodic evolution in the data. A major advantage of our method is that a critical parameter like rank of the model is automatically tuned using the automatic relevance determination (ARD) approach, unlike most matrix/tensor completion techniques. We also propose a robust version of the above formulation, which improves the performance of imputation in the presence of outliers.
We evaluate the proposed Variational Bayesian Filtering with Subspace Information (VBFSI) method to impute matrices in real-world traffic and air pollution data. Simulation results demonstrate that the proposed method outperforms the recent state-of-the-art methods and provides a sufficiently accurate imputation for different sampling rates. In particular, we demonstrate that fusing the subspace evolution over days can improve the imputation performance with even 15\% of the data sampling.

\end{abstract}

\section{Introduction}
Copious amounts of sensors are deployed in major cities for sensing the spatio-temporal variation in various urban environment characteristics like air pollution, traffic speed, etc.  Hitherto, the most common way of sensing data across a city is to deploy many stationary sensors or monitors across the city. Another way of sensing the spatio-temporal signals is to use drive-by sensing or moving sensors that sample various parts of a region at different time instances \cite{intro_drive1,hasenfratz2015deriving}. Devices that measure air quality, traffic speed, etc can be mounted on the moving vehicles to sample the data across different locations and time stamps based on the movement of the vehicle. Since one sensor can be used to sample many locations at different timestamps, the actual number of moving sensors needed is just a fraction of the total number of static sensors needed to get the same spatial coverage.
\par 

Data acquired from both static and moving sensors contains missing data due to sensor malfunction, irregularity in sensor measurements, etc.  Additionally, the drive-by sensing scheme uses relatively fewer sensors resulting in high data ``gaps" in both spatial and temporal dimensions \cite{intro_drive1}. This motivates the problem of extreme matrix completion, where the percentage of data sampled may be as low as $10\%$. Thereby, a natural question to ask is: how to fill the high percentage missing entries within a reasonable error range? Can we leverage additional periodic information in the matrix completion framework to estimate the high percentage of missing entries. Also, in addition to the missing entries, sensor measurements can be contaminated with outliers emerging from the sensor malfunctioning, communication errors, or impulse noises. The occurrence of outliers in the measurements can further degrade the performance of data imputation. However, unlike the missing entries, the location and the value of outliers are unknown, which makes the problem more challenging. Therefore, the other question to ask is: how to estimate the missing data while detecting the noisy outliers?  
 The answer to the above questions lies in exploiting the underlying structure available in the data. For instance, both the air pollution \cite{sampson2011pragmatic} and the traffic \cite{min2011real} data exhibit joint spatial and temporal correlation as well as periodicity in daily patterns, thereby generating redundancy that can be potentially exploited by performing an intelligent spatio-temporal extrapolation.   \par

\subsection{ Extreme Spatio Temporal Matrix Completion}
Data collected from the sensors in transportation and environment, etc., have spatial variability and follows slowly varying pattern over time. One way to exploit this spatiotemporal correlation is to represent the data in the form of a matrix, where one axis denotes the spatial variability, and the other axis indicates the temporal evolution.  
We motivate the problem of Extreme spatio-temporal matrix completion as follows. For a given day $d$, the data is represented in the form of a matrix $\X \in \mathcal{R}^{n \times t}$  where  $n$ and  $t$ are the numbers of spatial locations and time slots, respectively. The elements in this matrix are missing due to the moving sensor paradigm or sensor malfunctioning. The data will be highly sparse in the case of the moving sensor paradigm, where the sampling percentage can be as low as 10\%. Our goal is to estimate the missing entries in the matrix $\X$. 

One way of imputation is to exploit the low-rank structure available in the data. The data $\X$ is low rank because of the spatial correlation in the locations and the temporal correlation in the time slots \cite{asif2016matrix}. Further, the data represented in the matrix $\X$ is a time-series data. Enforcing only the low-rank structure in the data does not take into account the temporal variation (generally slow variation) of the data in a given location \cite{compare1}. We incorporate a state-space model to capture the temporal evolution in the data. Combining the low rankness and the temporal evolution estimates the data reasonably well in the scenario of low missing data \cite{compare6}. However, the performance deteriorates significantly for Extreme Matrix Completion when a higher percentage of data is missing. To estimate the entries in the case of Extreme Matrix completion, we propose to exploit the periodic pattern of the spatiotemporal data by reliable prior subspace information. Exploiting reliable prior subspace information can reduce the sample complexity of matrix completion and improve the imputation performance \cite{prove1}. We update the subspace on the day $d$ using the prior subspace estimated of the day $d-1$ to capture the periodicity in the data. Incorporating this prior subspace can improve the performance for an extreme case of sparsely sampled data. 

\subsection{Our contribution and Approach}
\begin{figure}[h]
	\centering
	\includegraphics[width=0.5\textwidth]{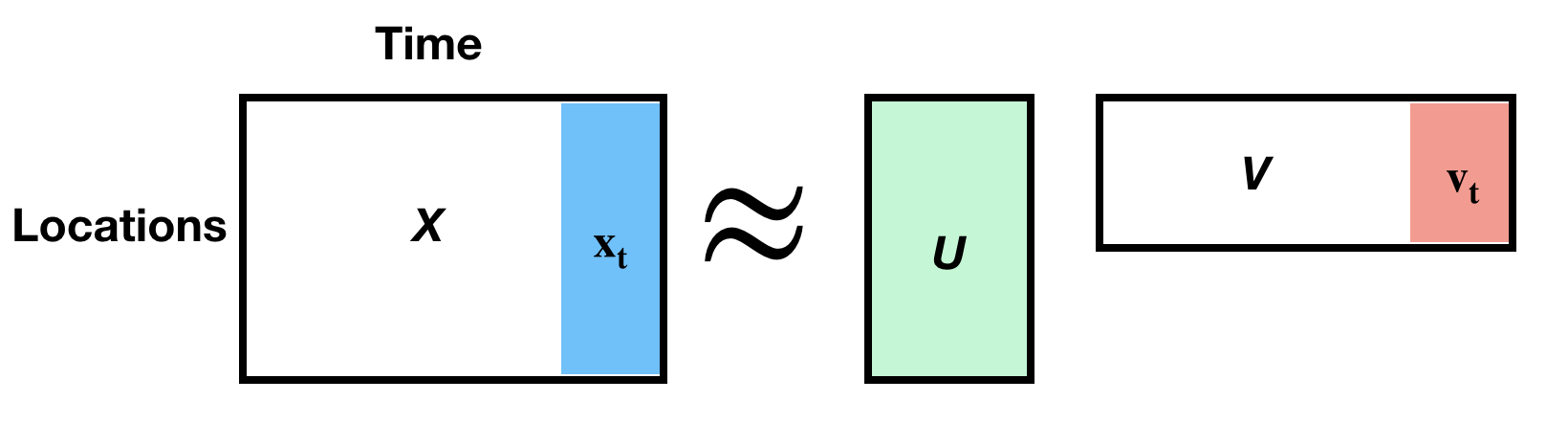}
	\centering
	\caption{ Low rank Matrix completion }
	\label{fig1}
\end{figure}

In this paper, we propose a Variational Bayesian Spatiotemporal matrix completion to estimate the data even in the presence of extreme matrix completion where the percentage of data sampled may be as low as $10\%$. We observe that even with a fraction of observed data, we can estimate the remaining measurements with reasonable accuracy. This work exploits the spatiotemporal and periodic pattern in the measurements for extreme matrix completion with reasonable accuracy. Firstly, we enforce a low-rank structure to the spatiotemporal data as shown in Fig \ref{fig1}. Secondly, we enforce the state-space model on the temporal embeddings $v_t$ to capture the temporal evolution in the data. Thirdly, we enforce the subspace estimate  $\mathbf{U}$of the matrix to be close to previously learned subspace distribution $\mathbf{\hat{U}}$ using the Mahalanobis distance. Exploiting the prior subspace over days results in a considerable reduction of the number of measurements required to estimate the matrix \cite{prove1,prove2} thereby boosting the performance in case of a low sampling rate.

We use the Variational Bayes approach to update the parameters in an iterative fashion. In our work, the subspace distribution is chosen to allow automatic relevance determination (ARD), and unlike the matrix or tensor completion methods, the algorithm parameters such as rank, noise powers need not be specified or tuned. We compare the performance of the proposed algorithm with various state-of-the-art algorithms on many real-world traffic and air quality datasets. The result shows that modeling the subspace evolution leads to improvement in performance even when a small percentage of random measurements are available for the purpose of imputation. A likely impact of our method is that cities with a low sensing budget can perform random drive-by sampling of the urban environment, and the suggested matrix completion framework can provide a reasonably accurate imputation leading to better decision making. \par 

Our main contribution can be summarized as follows.
\begin{itemize}
	\item  We propose Variational Bayesian Filtering with Subspace Information (VBFSI), a novel matrix completion framework that simultaneously models the low rank nature of the data, temporal evolution through a state space model, and periodicity through subspace tracking. 
	\item We experimentally showed that incorporating the prior subspace over days can improve imputation performance for extreme spatio temporal matrix completion i.e. for low data sampling.
	\item Critical parameter, like the rank of the model, is automatically tuned using the automatic relevance determination (ARD) approach. 
	\item  We also propose a Robust version of the VBFSI algorithm for imputing the data in the presence of outliers. 
	\item We conduct comprehensive experiments on real-world spatio-temporal datasets that show the efficacy of the proposed method over other state-of-the-art imputation methods. 
	
\end{itemize}

\subsection{Related work }
\begin{itemize}
	\item 	Big data matrices can be approximated as low-rank matrices \cite{udell2019big}. Matrix completion is used to exploit the low-rank structure in the data to impute the missing data  \cite{asif2016matrix,sparsebayesian}. Robust PCA is used for matrix completion in the presence of outliers by incorporating a sparse outlier matrix \cite{compare7,compare8}. The traditional matrix completion framework is not applicable for time series data imputation, as it does not take into account the ordering among the temporal embeddings \cite{compare1}. 
	\item Autoregressive model can model these temporal embeddings and in turn capture the temporal evolution in the time series data\cite{varst,compare1,compare6}. However, these models fail to capture the prior subspace information that can be exploited to capture periodicity in the time series data and only evaluated for lower percentage of missing data. 
	\item Exploiting the 3-way pattern in the data using tensor completion-based frameworks can incorporate the periodic pattern in the data and, in turn, improve the imputation performance \cite{asif2016matrix,compare3,ten1}. These method works better than the traditional matrix completion framework. However, the temporal evolution and subspace evolution is not modeled in the traditional tensor completion frameworks. We propose to incorporate the periodicity in the matrix completion framework while modeling the temporal and subspace evolution over days.
	
	\item Variational Bayesian approaches are proposed for matrix/tensor completion and robust principal component analysis by modeling the matrix/tensor as low rank \cite{sparsebayesian,var1,compare3,compare8,asif2016matrix}. A state-space model to capture the temporal evolution is also proposed in \cite{varst,compare6}. However, these approaches do not explicitly model the evolution of the subspace to capture the periodicity in the data.  
	\item The proposed matrix and tensor completion methods are not evaluated for extreme data missing problem and missing data estimation problem in the presence of outliers.

\end{itemize}

The paper is organized as follows: Section 2 presents the Variational Bayesian Filtering with Subspace Information. Section 3 presents the Robust Variational Bayesian Filtering with Subspace Information. Results and findings for air quality estimation and traffic estimation are discussed in section 4 followed by conclusion in section 5.

\subsection{Notations}
Tensor is denoted by $\pmb{X}$, Matrix is denoted by $\X$, its transpose is denoted by $\X^T$. For a given day $d$ the matrix is denoted by $\X^d$, we represent it as $\X$ throughout the paper. For the rest of the days matrix is represented by $\X^k$ where $k=\{1,2 \dots d-1\}$. The $(i,j)$-th element of a matrix $\X$ is denoted by $X_{i,j}$, the $i$-th row by $\x_{i\dott}$ and the $i$-th column by $\x_{i}$.  The multivariate Gaussian probability density function (pdf) with mean vector $\mub$ and covariance matrix $\Sig$ evaluated at $\x \in \Rn^n$ is denoted by $\N(\x \mid \mub,\Sig)$.\par


\section{Variational Bayesian Filtering with Subspace Information (VBFSI)}
\label{model}
Let  $\X  \in \Rn^{n \times t}$ be the data matrix for a day, where $n$ and $t$ denotes the number of spatial locations and time stamps respectively. 
The low rankness in the data can be imposed using the equation 
\begin{align}
\L_1= \min_{\U,\V}|| \mathbf{P}_{\Omega}  (\X-\U \V^{T})||_{{F}} \label{m1}
\end{align}
where $\U \in \mathcal{R}^{n \times r}$ and $\V \in \mathcal{R}^{t \times r}$ and $r=$rank$(\X) << min(n,t)$ implying the low rankness in the data. For sampling percentage of $p$, let $\Omega$  denotes the sampled data containing $p\times n\times t$ samples. $\mathbf{P}_{\Omega} $ is the indicator matrix where ${P}_{ij}=1 \,\, \forall (i,j) \in \Omega$.\\
To capture the temporal evolution in the data, we can regularize the columns of $\V$ to follow an autoregressive model.
\begin{align}
\mathcal{R}(\V)= \sum_{i=1}^{t}||\v_{i}-\F \v_{i-1}|| \label{m2}
\end{align}
Further, to capture the periodicity over days the subspace evolution can be modeled as:
\begin{align}
\mathcal{R}(\U)=\eta \sum_{i=1}^{n}(\u_{i}-\u_{i}^{d-1})^T(\Xib_i^{\U^{d-1}})^{-1} (\u_{i}-\u_{i}^{d-1}) \label{m3}
\end{align}
where $\mathcal{R}(\U)$ corresponds to the Mahalanobis distance between each row vector of $\U^{d-1}$ (subspace estimate of previous day) and $\U$ (current subspace estimate for a given day $d$). $\Xib^{\U^{d-1}}$ denotes the covariance matrix of $\U^{d-1}$. Here $\eta$ controls the effect of prior subspace ($\U^{d-1}$, $\Xib_i^{d-1}$) in the estimation of $\U$.

\subsection{Bayesian Model}
In this section, we will obtain a Bayesian framework for spatio temporal matrix completion. The optimization formulation in \eqref{m1} is equivalent to minimizing the negative log  likelihood function.
\begin{equation}
\L_1= \min_{\U,\V} (-\ln \p(\X_{\Omega}\mid\U,\V,\beta))
\end{equation}

where likelihood function on the entries of  $\X_\Omega$ can be defined as:
\begin{equation}
 \p(\X_{\Omega}\mid\U,\V,\beta)= \prod_{(i,j) \in \Omega }\mathcal{N} ( X_{ij}\mid \u_{i}. \v_{j}.^{T} ,\beta^{-1} )
\end{equation}
here $\beta$ is the noise precision. The prior on the noise is assumed to be non informative Jeffrey's prior. 
\begin{equation}
\p(\beta)=\beta^{-1}
\end{equation}

Regularization on the columns of $\U$ defined in \eqref{m3} can be incorporated by initializing  a prior on the columns of  $\U$. 
\begin{equation}
\p(\A \mid \gam) = \prod_{i=1}^r \N(\a_i \mid \a_i^{d-1}, \gamma_i^{-1}\I_n) \label{pa}
\end{equation}
Columns of  $\U$ are enforced with a sparsity profile using precison $\gamma_{i}$ to automate the rank. 
When $\gamma_{i}$ are driven to a large value then the column mean will be $\a_i^{d-1}$, and we prune these columns and in turn reducing the rank thereby modeling the low rank in the bayesian framework. As the columns with the high value of gamma are too simple to generate any data, whereas the columns with a low value of gamma are more powerful and can generate a greater variety of data \cite{mackay1992bayesian}. This way of determining the rank on the go is referred to the as the Automatic Rank Determination \cite{sparsebayesian}. 
Further the  autoregressive regularization in \eqref{m2} can be modeled as 
\begin{equation}
\p(\B \mid \J) = \N(\b_1; \mub_1, \Lam_1 ) \prod_{\tau = 2}^t \N(\b_\tau \mid \J\b_{\tau-1}, \I_r) \label{pb}
\end{equation}
$\J$ is assigned multivariate Gaussian priors with column-specific precisions $\upsilon$.
\begin{equation}
\p(\J \mid \up) = \prod_{i=1}^r \N(\j_i \mid 0, \upsilon_i^{-1}\I_r) \label{pj}
\end{equation}
Precision variables  $\gam$ and $\up$ are selected to have non-informative Jeffrey's priors
\begin{align}
\p(\gamma_i) = \frac{1}{\gamma_i},\,\, & \p(\upsilon_i) = \frac{1}{\upsilon_i}
\end{align}
The overall joint distribution for spatio-temporal matrix completion can be expressed as
\begin{align}
\p(\Y_\Omega,\U,\V,\F,\beta,\gam,\up)= \p(\Y_\Omega | \A, \B, \beta)\p(\A | \gam) &\nonumber\\
&\hspace{-4.5cm} \times \p(\B | \J) \p(\J | \up)\p(\beta)\p(\up)\p(\gam)
\end{align}

The Full Bayesian graphical model for spatio-temporal matrix completion is shown in Fig \ref{fig:emc_algo_1}. 
\def \U {\mathbf{U}}

\def \v {\mathbf{v}}
\def \Xib {\boldsymbol{\Xi}}
\def \up {\boldsymbol{\upsilon}}
\begin{figure}[ht]
  \begin{center}
    \begin{tabular}{cc}
     	\begin{tikzpicture}
			\node[obs]          (y)   {$\x_{\tau}$}; %

			\node[latent, left=1.5cm of y]         (a)   {$\U$}; %
			\node[latent, left=1cm of a,above]         (a1)   {$\U^{d-1}$}; 
			\node[latent, left=1cm of a,below]         (a2)   {$\Xib_i^{\U_{d-1}}$}; 
			
			\node[latent, below =0.5 of y]  (b)   {$\mathbf{v}_{\tau}$}; %
			
			\node[latent, left =0.5cm of b] (b1)  {$\mathbf{v}_{\tau-1}$} ; %
			\node[latent, above=0.4cm of a] (r)  {$\gam$} ; %
			\node[latent, below=0.9 of b] (j)   {$\mathbf{F}$}; %
			\node[latent, left =0.5cm of j] (v)  {$\up$} ; %
			
			\node[latent, right=1cm of y]         (be)   {$\beta$}; %
		
			\edge {be,b,a} {y} ; 
			\edge {r,a1,a2} {a} ; 
			\edge {b1,j} {b} ;

			\edge {v} {j} ; 
			
			\plate {yx} {(y)(b)(b1)}{$t$} ;
			\plate {yz} {(r)(b)(j)(be)}{$n$} ;
			\end{tikzpicture}
    \end{tabular}
  \end{center}
	\caption{Variational Bayesian Filtering with Subspace Information}
	\label{fig:emc_algo_1}
\end{figure}
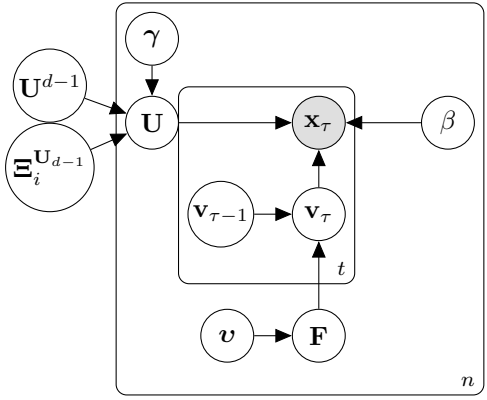

\subsection{Variational Bayesian Inference}

We utilize the mean-field approximation, wherein the posterior distribution of parameters $\th:= \{\U,\V,\J,\beta,\gam, \up\}$ is factorized into a set of conditionally independent components. It is expressive as it captures the marginal density of the parameters. The main advantage of this assumption is that the optimization takes the form of coordinate ascent where the posterior distribution of each parameter can be found by taking expectation of all the other parameters in an iterative manner. The posterior distribution of parameters is factorized as:
\begin{align}\label{mf}
\p(\th \mid \y_\Omega)= q_{\A}(\A)q_{\B}(\B)q_{\J}(\J)q_{\up}(\up)q_{\beta}(\beta)q_{\gam}(\gam).
\end{align}

The posterior distribution of all the parameters is determined by minimizing the Kullback--Leibler divergence of $\p(\th|\y_\Omega)$ from $q(\th)$, usually via an alternating minimization approach  \cite{bishop2006pattern}.\par
We use variational inference to estimate the  posterior distribution of $q_\A$, $q_\V$, $q_\J$, $q_{\up}$, $q_\beta$, and $q_{\gam}$ for sampled data $\Z=\mathbf{P}_\Omega(\X)$. The updates for posterior distribution of parameters are similar to the updates derived in \cite{sparsebayesian, compare6}.\\
The posterior distribution for a row of $\A$ is given by 
\begin{align}
q_{\u_{i}} &= \N({\u_{i}} \mid \mub_i^{\U}, \Xib_i^{\U}) 
\end{align}
The updates of mean and covariance of $\U$ are derived as 
\begin{align}
(\Xib^\U_i)^{-1}=  \hat{\gamma}_i\I_{r} + \hat{\beta}\sum_{\tau| (i,\tau) \in\Omega}(\mub_\tau^\V(\mub_\tau^\V)^T+\Xib_{\tau,\tau}^\V) &\nonumber\\
&\hspace{-5cm}+\eta( \Xib_{i}^{\A_{d-1}})^{-1} 
\end{align}
\begin{align}
\mub^\A_i =\Xib^\A_i( \hat{\beta}\sum_{\tau| (i,\tau) \in\Omega} \mub_\tau^\V Z_{i\tau}+\eta (\Xib_{i}^{\A_{d-1}})^{-1} \mub_i^{\A_{d-1}})
\end{align}

The mean and covariance for the Posterior Distribution of  $\V$ are as follows:
\begin{align}
q_{\V}(\V) &= \N(\vec{\V} \mid \mub^{\V}, \Xib^{\V})
\end{align}
\begin{align}
\mub^\V &= \Xib^\V\begin{bmatrix} \hat{\beta}\sum_{i|(i,1)\in\Omega}Z_{i,1}\mub^\A_i  + \Lam_1^{-1}\mub_1\\
\hat{\beta}\sum_{i|(i,2)\in\Omega}Z_{i,2}\mub^\A_i \\
\vdots\\
\hat{\beta}\sum_{i|(i,t)\in\Omega}Z_{i,t}\mub^\A_i
\end{bmatrix}\label{mub}
\end{align}

\begin{align}
\left[\Xib^{\V}\right]^{-1} &=  \hat{\beta}\Diag{\Xib^\A_{(1)}, \ldots, \Xib^\A_{(t)}}  + \nonumber\\
&+ \begin{bmatrix} \Lam_1^{-1} & -\hat{\J} & \ldots &0\\
-\hat{\J} & \I_r + \Sig^\J & -\hat{\J}  & \ldots \\
\vdots & \vdots &&\vdots \\
\ldots & 0 & -\hat{\J} &  \I_r 
\end{bmatrix} \label{xib}
\end{align}
The direct inversion of the dense matrix $\Xib^\V$ would be computationally costly. The  block-tridiagonal structure $(\Xib^\V)^{-1}$  can be exploited to carry out the updates for $\Xib^\V$ in an efficient manner using LDL decomposition \cite{varst,compare6}. 

The updates of the posterior distribution of $\F$ are given by
\begin{align}
q_{\j_{i}} &= \N({\j_i} \mid \mub_i^{\J}, \Xib_i^{\J})
\end{align}
\begin{subequations}
	\begin{align}
	\mub^\J_i &= [\Xib^\J_i(\mub_\tau(\mub_{\tau-1})^T+\Xib_{\tau,\tau-1}^\V)]_{ i} \\
	({\Xib^\J_i})^{-1} &= \Diag{\hat{\up}} + \sum_{\tau=1}^{t-1}(\mub_\tau(\mub_{\tau-1})^T+\Xib_{\tau,\tau-1}^\V)\label{15b}
	\end{align}
\end{subequations}
The posterior distribution for hyperparameters $\{\beta, \gam, \v\}$ are given by

\begin{subequations}
	\begin{align}
	q_{\beta}(\beta) &= \text{Ga}(\beta; a^\beta, b^\beta) \\
	q_{\gamma_i}(\gamma_i) &= \text{Ga}(\gamma_i; a_i^\gamma, b_i^\gamma) \\
	q_{\upsilon_i}(\upsilon_i) &= \text{Ga}(\upsilon_i; a_i^\upsilon, b_i^\upsilon) 
	\end{align}
\end{subequations}
where Ga$(x, a, b)$ denotes the Gamma pdf with parameters $a$ and $b$. The updates for  $\{\beta, \gam, \v\}$ are given by
\begin{subequations}\label{upga}
	\begin{align}
	\hat{\upsilon}_i &= \frac{r}{\sum_{k=1}^r\left([\mub^\J_k]^2_i + [\Xib^\J_k]_{ii}\right)} \label{10a} \\
	\hat{\gamma}_i &= \frac{n+t}{\sum_{k=1}^n\left([\mub^\A_k]^2_i + [\Sig^\A_k]_{ii}\right)+{\sum_{k=1}^t\left([\mub^\V_k]^2_i + [\Sig^\V_k]_{ii}\right)}}\label{10b} \\
	\hat{\beta} &=\frac{p\,n\,t}{\parallel  \mathbf{\Z}- P_{\Omega}(\A\,\V\,^T)\parallel^2_{F}}
	\end{align}
\end{subequations}
We update the mean, covariance of $\U,\V,\F$ and the hyperparameters $\gam,\up,\beta$ iteratively as shown in Algorithm \ref{Algo 1}
\begin{algorithm}
	\caption{VBSFI}
	\label{alg:algorithm}
	\textbf{Input}: $\Xib^{\A_{d-1}},\mub^{\A_{d-1}},\mathbf{P}_\Omega(\X)$\\
	\textbf{Initialization}:$\gam,\beta,\up,\Xib^\A,\mub^\A, \Xib^\V,\mub^\V,\Xib^\J,\mub^\J,\Z $
	
	\begin{algorithmic}[1] 
		
		\WHILE{$X_{conv}< 10^{-5}$}
		
		\STATE 	$\mathbf{X_{old}}= \hat{\Y}$\\
		Compute $\V,\F,\up,\beta$ using (21, 22, 24, 26a, 26c)\\
		Compute $\U,\gam,\beta$ using (18, 19, 26b, 26c)\\
		$\hat{\Y}=\mub^\A(\mub^\V)^T$\\
		$X_{conv}=\frac{\norm{\hat{\Y}-\mathbf{X_{old}}}_F}{\norm{\mathbf{X_{old}}}_F}$
		
		\ENDWHILE
		\STATE 
		\textbf{Output}: $\hat{\Y}$
	\end{algorithmic}
	\label{Algo 1}
\end{algorithm}

\section{Robust Variational Bayesian Filtering with Subspace Information (RVBFSI)}
In this section we consider the robust version of the Variational Bayesian Filtering with Subspace Information. RVBFSI estimates the missing data while detecting the noisy outliers. 
For robust matrix completion, we model $\X=\U \V^T+\E+\No$, where $\E$ denotes the sparse outlier matrix and $\No$ is the dense error matrix. The low rankness defined in the Eq. \ref{m1} is modified to incorporate the sparse outlier matrix into the framework as
\begin{align}
\L_1= \min_{\U,\V}|| \mathbf{P}_{\Omega}  (\X-\U \V^{T}-\E)||_{{F}} \label{em1}
\end{align}
The regularization on $\U$ and $\V$ to model the temporal and subspace evolution follows the Eq. \ref{m2}-\ref{m3}. 
\subsection{Bayesian Model}
The conditional distribution of generating the entries of $\X_{\Omega}$ can be defined as  
\begin{align}
\ \p(\X_{\Omega}\mid\U,\V,\E,\beta)= \prod_{(i,j) \in \Omega }\mathcal{N} (X_{ij}\mid (\u_{i}. \v_{j}.^{T} + E_{ij}),\beta^{-1} )
\end{align}
Columns of $\U$ ,$\V$ and $\F$  and precision variables $\gam, \beta, \upsilon$ follows same the prior distribution defined in (6-10).\\
Each entry of sparse outlier matrix $E_{ij}$ is assigned a precision $\alpha_{ij}$.
\begin{align}
\p(\E|\al) &=\prod_{i|((i,j) \in \Omega)}\prod_{j|((i,j) \in \Omega)} \N(E_{ij} \mid 0, \alpha_{ij}^{-1}) 
\end{align}
where the $\alpha_{ij}$ have the non informative prior
\begin{align}
\p(\alpha_{ij}) &= \frac{1}{\alpha_{ij}}
\end{align}
This works similar to the ARD where instead of column of the matrix, each entry of the matrix is assigned with a precision. Whenever $\alpha_{i,j}$ is driven to a large value , the $E_{i,j} \to 0$ thereby enforcing sparsity.
The overall joint distribution for Robust Spatio-Temporal Matrix Completion is expressed as
\begin{align}
\p(\Y_\Omega,\U,\V,\F,\E,\beta,\gam,\upsilon,\alpha) = \p(\Y_\Omega | \A, \B, \E,\beta)\p(\A | \gam)&\nonumber\\
&\hspace{-6.3cm}\times \p(\B | \J) \p(\J | \up)\p(\E | \al)\p(\beta)\p(\up)\p(\gam)
\end{align}
The full bayesian model for the Robust Spatio-Temporal Matrix Completion is depicted in \ref{fig:emc_algo_2}. 
\subsection{Variational Bayesian Inference}

Approximate posterior distrubution of parameters $\th_R:= \{\U,\V,\J,\E,\beta,\gam,\up,\al\}$ are derived using Variational Inference.

We utilize the mean-field approximation, wherein the posterior distribution of parameters $\th_R$ is factorized as:
\[q_{\A}(\A)q_{\B}(\B)q_{\J}(\J)q_{\E}(\E)q_{\up}(\up)q_{\beta}(\beta)q_{\gam}(\gam)q_{\al}(\al).\]

 The posterior distribution of $q_\A$, $q_\V$, $q_\J$, $q_{\up}$, $q_\beta$, and $q_{\gam}$ takes the same form for $\Z=\mathbf{P}_\Omega(\X-\E)$ as shown in (17-26c). The posterior distribution for $\E$ take the following form $\forall (i,j) \in \Omega$.
\begin{align}
q({E_{ij}}) = \N({E_{ij} \mid \mu_{ij}^{\E}, \Xi_{ij}^{\E}})
\end{align}
\begin{align}
\Xib_{i,j}^{\E}=\frac{1}{\hat{\beta}+\hat{\alpha_{i,j}}}
\end{align}
\begin{align}
\mub_{i,j}^{\E}=\hat{\beta}\, \Xib_{i,j}^{\E}(X_{i,j}-\mub_{i.}^A(\mub_{j.}^B)^T)
\end{align}

\begin{align}
\hat{\alpha}_{i,j}^{new}=\frac{1-\hat{\alpha}_{i,j}^{old}  \,\Xib_{i,j}^\E}{(\mub_{i,j}^\E)^2}
\end{align}
$\hat{\alpha}_{i,j}^{new}$ is the fixed-point update for $\al$. This is used in the sparse bayesian learning that leads to much faster convergence and enhanced sparsity \cite{sparsebayesian,sparsebayesian2}.
For robust estimation of entries in the presence of outliers, we update the mean, covariance of $\U,\V,\F,\E$ and the hyperparameters $\gam,\up,\beta,\al$ iteratively as shown in Algorithm \ref{Algo2}.
\def \U {\mathbf{U}}

\def \v {\mathbf{v}}
\def \Xib {\boldsymbol{\Xi}}
\def \up {\boldsymbol{\upsilon}}
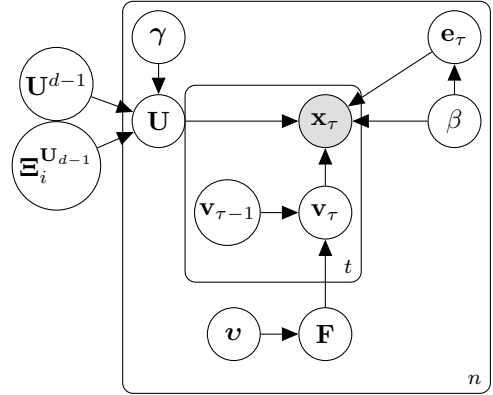
\begin{figure}[ht]
  \begin{center}
    \begin{tabular}{cc}
   	\begin{tikzpicture}
			\node[obs]          (y)   {$\x_{\tau}$}; %

			\node[latent, left=1.5cm of y]         (a)   {$\U$}; %
			\node[latent, left=1cm of a,above]         (a1)   {$\U^{d-1}$}; 
			\node[latent, left=1cm of a,below]         (a2)   {$\Xib_i^{\U_{d-1}}$}; 
			
			\node[latent, below =0.5 of y]  (b)   {$\mathbf{v}_{\tau}$}; %
			
			\node[latent, left =0.5cm of b] (b1)  {$\mathbf{v}_{\tau-1}$} ; %
			\node[latent, above=0.4cm of a] (r)  {$\gam$} ; %
			\node[latent, below=0.9 of b] (j)   {$\mathbf{F}$}; %
			\node[latent, left =0.5cm of j] (v)  {$\up$} ; %
			
			\node[latent, right=1cm of y]         (be)   {$\beta$}; %
			\node[latent, above=0.4cm of be]         (e1)   {$\e_\tau$}; 
			
			\edge {be,b,a,e1} {y} ; 
			\edge {r,a1,a2} {a} ; 
			\edge {b1,j} {b} ; 
			\edge {be} {e1} ; 
			
			\edge {v} {j} ; 
			
			\plate {yx} {(y)(b)(b1)}{$t$} ;
			\plate {yz} {(r)(b)(j)(be)}{$n$} ;
			\end{tikzpicture}
    \end{tabular}
  \end{center}
	\caption{Robust Variational Bayesian Filtering with Subspace Information}
	\label{fig:emc_algo_2}
\end{figure}

\begin{algorithm}
	\caption{RVBSFI}
	\label{alg:algorithm}
	\textbf{Input}: $\Xib^{\A_{d-1}},\mub^{\A_{d-1}},\mathbf{P}_\Omega(\X)$\\
	\textbf{Initialization}:$\gam,\beta,\up,\Xib^\A,\mub^\A, \Xib^\V,\mub^\V,\Xib^\J,\mub^\J,\mub^\E, \Xib^\E,\\\al, \Z=\mathbf{P}_\Omega(\X-\E) $
	
	\begin{algorithmic}[1] 
		
		\WHILE{$X_{conv}< 10^{-5}$}
		
		\STATE 	$\mathbf{X_{old}}= \hat{\Y}$\\
		Compute $\V,\F,\up,\beta$  using (21, 22, 24, 26a, 26c)\\
		Compute $\U,\gam,\E,\al, \beta$ (18, 19, 26b, 26c)\\\
		Compute $\E,\al, \beta$ using (28-29, 26c)
		
		$\hat{\Y}=\mub^\A(\mub^\V)^T$\\
		$X_{conv}=\frac{\norm{\hat{\Y}-\mathbf{X_{old}}}_F}{\norm{\mathbf{X_{old}}}_F}$
		
		\ENDWHILE
		\STATE 
		\textbf{Output}: $\hat{\Y}$
	\end{algorithmic}
	\label{Algo2}
\end{algorithm}

\section{Experimentation}
In this section we will evaluate the performance of VBFSI on various spatiotemporal datasets against the recent state of the art imputation methods. We further compare the performance of RVBFSI in the presence of artificially corrupted outliers.
We will answer the following research questions.\par
  \textbf{RQ1}: How does our proposed VBFSI compare to recent state-of-the-art matrix/tensor imputation methods for spatio-temporal datasets?\par 
 \textbf{RQ2}: What is the effect of $\eta$ on the performance of VBFSI? \par
  \textbf{RQ3}: What is the effect of Outlier on RVBFSI , VBFSI and other recent state-of-the-art matrix/tensor imputation methods ?
\subsection{Experiment Setting}
\subsubsection{Datasets}
We used traffic speed and air quality (PM 2.5) data for performance evaluation. 
\begin{itemize}
	
	\item Data (DT): Delhi traffic speed data \cite{compare6}. This data contains traffic speed data of 519 road segments over 60 days with a sampling resolution of 15 min from 7 am to 11 pm in Delhi, India. The data can be organized as a tensor with dimensions $\pmb{R}^{519\times67\times60}$.
	\item Data (GT): Guangzhou urban traffic speed data \cite{dataset1}. This data contains traffic speed data of 214 road segments over 61 days with a sampling resolution of 10 mins in Guangzhou, China. The data can be organized as a tensor with dimensions $\pmb{R}^{214\times144\times61}$. 
	\item Data (PT): Pems traffic speed data \cite{dataset2}. This data contains traffic speed data of 228 road segments over 44 days with a sampling resolution of 5 mins in California. We process the data for a sampling resolution of 30 mins. The data can be organized as a tensor with dimensions $\pmb{R}^{228\times48\times44}$. 
	\item Data (CA): China Air Quality data \cite{dataset3}. This data contains the AQI  data collected in the cities near Beijing and Guangzhou in China. We pre-process the data and extract the PM2.5 AQI data for 313 locations and 60 days with a sampling resolution of 1 hr. The data can be organized as a tensor with dimensions $\pmb{R}^{313\times24\times60}$. 
\end{itemize}

\subsubsection{Parameters Setting}
\begin{figure}[ht!]
    \centering
    \begin{subfigure}{0.8\linewidth}
        {\label{fig:hyper1}\includegraphics[clip, trim=3.5cm 0.5cm 2.5cm 1cm,width=0.75\linewidth]{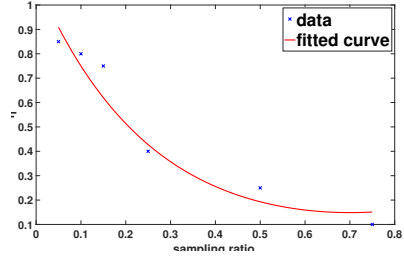}}
        \caption{$\eta$ vs $p$ performance for Data:DT }
    \end{subfigure}
    \begin{subfigure}{0.8\linewidth}
       {\label{fig:hyper2}\includegraphics[clip, trim=3.5cm 0.5cm 2.5cm 1cm,width=0.75\linewidth]{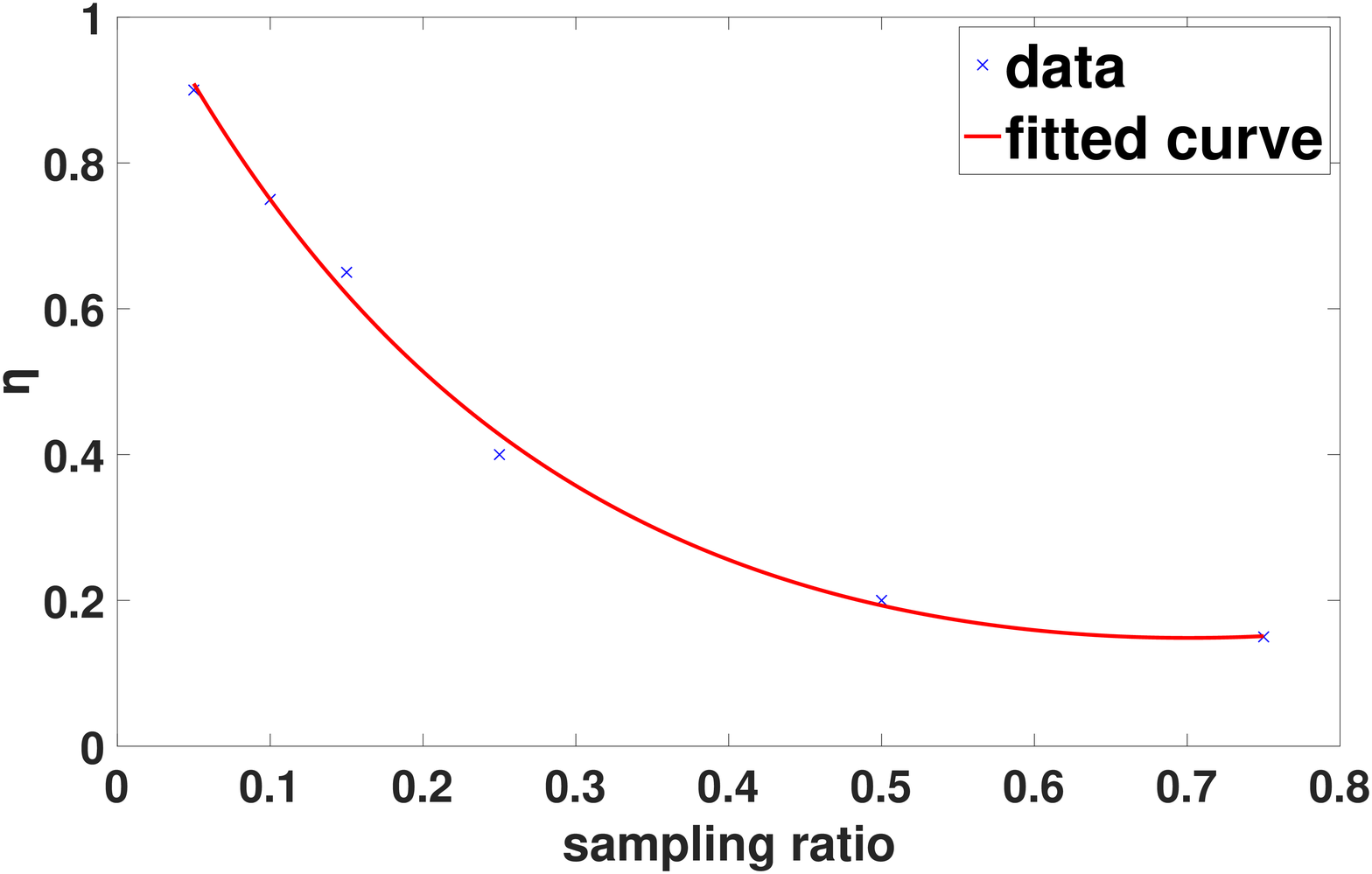}}
        \caption{$\eta$ vs $p$ performance for Data:CA}
    \end{subfigure}
  
    \caption{Hyperparameter setting for $\eta$. }
    \label{fig:hyper}
\end{figure}

The parameters we used in our experiment are described as follows:
\begin{itemize}
    \item We use EM algorithm to approximate the posteriors of all the model parameters and hyperparameters. Our method is characterized as a tuning parameter-free approach that can effectively avoid parameter selections. The top level hyperparameters including $a_\gamma$ , $b_\gamma$ are set to $10^6$ , resulting in a noninformative prior. Rank determination is automatic. We only tune the parameter$\eta$.
    \item We grid search the best $\eta$ for different sampling percentage. Then, we fit the exponential model for  $\eta$ vs.  $p$, as shown in Fig. \ref{fig:hyper}. We evaluate the algorithm for $\eta$[1,0.9, 0.75,0.5,0.25,0.1] for sampling percentage $p$ [0.05,0.1,0.15,0.25,0.5,0.75].We observe that for a higher sampling percentage, imputation performance decreases with an increase in $\eta$. After fitting the exponential model for traffic data (DT) $\eta $ is set as $\eta=1.09*exp(-3.87*p) + 0.00862*exp(3.76*p)$ whereas $\eta=1.282*exp(-11.18*p)+0.0289*exp(1.74*p)$ is set for air quality data (CA). We tune $\eta$ for data (DT) and generalize it for all other two traffic data (PT and GT).  \par
    \item The initial subspace $\U^0$ is calculated using the eight days average for all the datasets. Then we run the algorithm in an online fashion for the next 30 days for all the datasets. All the experiments are run on Matlab with the system configuration of 2.3 GHz and 8 GB RAM.

\end{itemize}

\begin{table*}[t]
	\centering
	\small
	
	\begin{tabular}{|l|l|l|l|l|l|l|l|l|}
		\hline
		& p \% & VBFSI & VBSF & VMC & BCPF & TRLRF & TRMF & BTMF \\ \hline
		Data:DT & 5\% & \bftab 0.156 / 4.387 & 0.782 / 22.03 & 0.998 / 28.18 & 0.164 / 4.6 & 0.901 / 25.45 & 0.183 / 5.146 & \underline{0.157 / 4.394} \\
		& 15\% & \bftab 0.135 / 3.796 & 0.162 / 4.552 & 0.97 / 27.39 & 0.147 / 4.137 & 0.682 / 19.25 & 0.151 / 4.24 & \underline{0.137 / 3.836} \\ 
		& 25\% &\underline{ 0.127 / 3.576} & 0.142 / 3.999 & 0.155 / 4.357 & \bftab0.126 / 3.544 & 0.415 / 11.72 & 0.135 / 3.801 & 0.129 / 3.613 \\
		& 50\% &\underline{ 0.117 / 3.289} & 0.119 / 3.344 & 0.131 / 3.687 & \bftab0.115 / 3.23 & 0.171 / 4.785 & 0.121 / 3.409 & 0.119 / 3.342 \\ 
		& 75\% & \underline{0.11 / 3.086} & 0.11 / 3.099 & 0.117 / 3.28 & \bftab0.109 / 3.076 & 0.13 / 3.642 & 0.117 / 3.262 & 0.115 / 3.224 \\ \hline
		Data:PT & 5\% & \bftab0.144 / 8.608 & 1 / 60.08 & 0.998 / 59.974 & 0.175 / 10.494 & 0.94 / 56.47 & 0.161 / 9.571 & \underline{0.151 / 9.084} \\ 
		& 15\% & \bftab0.111 / 6.625 & 0.179 / 10.7 & 0.974 / 58.52 & 0.147 / 8.836 & 0.807 / 48.49 & 0.139 / 8.264 & \underline{0.118 / 7.06} \\ 
		& 25\% &\bftab 0.101 / 6.026 & 0.147 / 8.82 & 0.152 / 9.13 & \underline{0.108 / 6.492} & 0.605 / 36.35 & 0.118 / 6.995 & 0.11 / 6.571 \\ 
		& 50\% & \bftab 0.084 / 5.056 & 0.097 / 5.79 & \underline{0.087 / 5.213} & 0.091 / 5.431 & 0.168 / 10.07 & 0.093 / 5.511 & 0.1 / 6.02 \\ 
		& 75\% & \underline{0.081 / 4.841} & 0.081 / 4.854 & \bftab0.069 / 4.135 & 0.081 / 4.848 & 0.097 / 5.833 & 0.083 / 4.951 & 0.097 / 5.81 \\ \hline
		Data:GT & 5\% & 0.159 / 6.384 & 1 / 40.31 & 0.993 / 40.03 & \underline{0.158 / 6.346} & 0.863 / 34.76 & 0.184 / 6.614 & \bftab0.131 / 5.244 \\ 
		& 15\% & \underline{0.121 / 4.854} & 0.148 / 5.91 & 0.382 / 15.33 & 0.138 / 5.541 & 0.492 / 19.8 & 0.162 / 5.845 & \bftab 0.11 / 4.43 \\
		& 25\% &\bftab 0.106 / 4.24 & 0.145 / 5.813 & 0.111 / 4.475 & 0.114 / 4.583 & 0.214 / 8.597 & 0.144 / 5.2 & \underline{0.103 / 4.126} \\ 
		& 50\% & \bftab0.088 / 3.547 & 0.112 / 4.501 & \underline{0.09 / 3.616} & 0.097 / 3.902 & 0.112 / 4.503 & 0.128 / 4.624 & 0.095 / 3.801 \\ 
		& 75\% & \bftab 0.079 / 3.189 & 0.1 / 4.027 & \underline{0.081 / 3.247} & 0.088 / 3.515 & 0.091 / 3.652 & 0.12 / 4.303 & 0.092 / 3.712 \\ \hline
		Data:CA & 5\% & 0.439 / 32.44 & 1 / 76.562 & 0.998 / 76.464 & 0.435 / 32.672 & 0.978 / 74.949 & 0.434 / 33.915 & \bftab0.414 / 31.431 \\ 
		& 15\% & 0.35 / 25.964 & 0.396 / 29.762 & 0.986 / 75.582 & \bftab0.341 / 25.471 & 0.936 / 72.046 & 0.369 / 28.735 & \underline{0.344 / 25.94}\\ 
		& 25\% & 0.304 / 22.578 & 0.308 / 23.05 & 0.679 / 50.661 & \bftab0.297 / 22.207 & 0.886 / 68.426 & 0.32 / 24.908 &\underline{0.293 / 22.045} \\ 
		& 50\% & \underline{0.222 / 16.466} & 0.23 / 17.172 & \bftab0.213 / 15.892 & 0.237 / 17.646 & 0.731 / 57.059 & 0.235 / 18.274 & 0.248 / 18.483 \\ 
		& 75\% & \underline{0.198 / 14.648} & 0.2 / 14.675 & \bftab0.171 / 12.636 & 0.209 / 15.603 & 0.472 / 36.816 & 0.197 / 15.249 & 0.223 / 16.556 \\ \hline
	\end{tabular}
	\caption{MRE/RMSE scores for data imputation. The best two results are bold and underlined respectively.}
	
	\label{table:table1}
\end{table*}

\subsubsection{Evaluation Metrics}
We use the Mean relative error (MRE), and root mean square error (RMSE) as evaluation metrics:
\[\small\text{MRE}= \frac{\norm{\X(\Omega')-\hat{\X(\Omega')}}}{\norm{\X(\Omega')}} \]
\[\small\text{RMSE}= \sqrt{\frac{1}{|\Omega'|}\sum_{(i,j\in \Omega') } (\hat{X_{ij}}-X_{ij})^2}\]
where $\Omega'$ represent the set of missing entries. 
\subsection{ Baseline Algorithms}
We compare our model with recent state of the art matrix and tensor imputation methods. 
\subsubsection{Matrix completion Frameworks}
\begin{itemize}
	\item \textbf{VBSF}: Variational Bayesian Subspace Filtering \cite{compare6}, 
	VBSF proposes an Variational Bayesian formulation to estimate low-rank matrices whose subspace evolves according to a state-space model.
	\item \textbf{VMC}: Variety-based Matrix Completion \cite{compare4}. VMC exploit low-complexity nonlinear structures in the data to estimate the matrix that can be possible high-rank. The high rank matrix becomes low-rank after mapping each column to a higher dimensional space.
\end{itemize}

\begin{itemize}
	\item \textbf{TRMF}: Temporal regularized matrix factorization \cite{compare1}. TRMF exploits the autoregressive structure among temporal embeddings $\v_t$.
	TRMF uses a set ($L$) containing the lag indices $l$ denoting a dependency between $t^{th}$ and $(t-  l)^{th}$ time points. We take the lag as \{1, 2,$T$\}, where $T$  denotes the number of time intervals in a day. We stack $d-8, d-7,\dots d$  data matrices to predict the samples for $\text{d}^{th}$ day, thereby incorporating the dependencies over days and week.
	\item \textbf{BTMF}: Bayesian Temporal Matrix Factorization \cite{compare2} is a bayesian extension of TRMF which outperforms TRMF and other imputation methods for traffic data. 
\end{itemize}
\subsubsection{Tensor Completion Frameworks}   
To evaluate the performance of the tensor completion algorithms with VBFSI we use $\pmb{X} \in \pmb{R}^{n\times t \times k }$, a three way tensor. For an effective comparison between matrix and tensor completion frameworks, $k$ is set as 7 \cite{asif2016matrix}. However, we set the $k=8$ to capture the weekly pattern, usually observed in traffic data. Also, we observe that the performance is better for $k=8$ as compared to $K=7$
\begin{itemize}
	\item \textbf{BCPF}:  Bayesian CP Factorization \cite{compare3}. BCPF is a bayesian tensor-based imputation method that incorporates a sparsity-inducing prior over multiple latent factors. BCPF is effective even for a higher percentage of missing data. 
	\item \textbf{TRLRF}: Tensor ring low-rank factors \cite{compare5} is an efficient and high-performance tensor completion algorithm based on TR(Tensor Ring) decomposition, which employed low-rank constraints on the TR latent space. TRLRF outperforms the state of the art tensor completion algorithm for synthetic and real-world data.
\end{itemize}

\subsubsection{Robust Imputation Frameworks}   
We compare RVBFSI with the following Robust imputation methods.
\begin{itemize}
\item \textbf{RVBSF}: Robust Variational Bayesian Subspace Filtering \cite{compare6}, 
	RVBSF proposes an robust variational Bayesian formulation to estimate low-rank matrices whose subspace evolves according to a state-space model in the presence of outlier.
	\item \textbf{Reg$\text{L}_1$}:  Regularized $\text{L}_1$ Augmented Lagrange Multiplier \cite{compare7} is proposed to approximate a low-rank data matrix in the presence of missing data and outliers.
	\item \textbf{BRTF}: Bayesian Robust Tensor Factorization \cite{compare8} uses variational bayesian approach for robust tensor factorization in the presence of  missing entries and outliers. 
\end{itemize}

\begin{table*}
	\centering
	\small
	
	\begin{tabular}{|l|l|l|l|l|l|l|l|l|}
		\hline
		& \multicolumn{4}{c|}{o=5\% } &  \multicolumn{4}{c|}{o=10\% } \\ \hline
		p \% & 10\% & 25\% & 50\% & 75\% & 10\% & 25\% & 50\% & 75\% \\ \hline
		RVBFSI & \bftab0.167 / 4.672 &\bftab 0.14 / 3.91 & \bftab0.126 / 3.544 &\bftab 0.119 / 3.337 & \bftab0.17 / 4.78 & \bftab0.14 / 3.925 &\bftab 0.128 / 3.573 & \bftab0.118 / 3.314 \\ \hline
		RVBSF & {0.196 / 5.527} & \underline{0.154 / 4.313} & \underline{0.132 / 3.696} & \underline{0.124 / 3.485} & 0.227 / 6.403 & \underline{0.16 / 4.492} & \underline{0.132 / 3.728} & \underline{0.124 / 3.487} \\ \hline
		VBFSI & \underline{0.188 / 5.277} & 0.177 / 4.972 & 0.17 / 4.778 & 0.158 / 4.427 & 0.208 / 5.86 & {0.194 / 5.443} & 0.184 / 5.173 & 0.175 / 4.899 \\ \hline
		VBSF & 0.305 / 8.583 & 0.201 / 5.656 & 0.177 / 4.98 & 0.166 / 4.65 & 0.391 / 11.01 & 0.237 / 6.65 & 0.198 / 5.561 & 0.184 / 5.16 \\ \hline
		VMC & 0.995 / 28.05 & 0.798 / 22.4 & 0.368 / 10.35 & 0.306 / 8.605 & 0.996 / 28.08 & 0.937 / 26.41 & 0.562 / 15.8 & 0.44 / 12.38 \\ \hline
		BCPF & 0.193 / 5.422 & {0.164 / 4.597} & 0.146 / 4.101 & 0.14 / 3.927 & \underline{0.205 / 5.763} & 0.179 / 5.017 & 0.155 / 4.363 & 0.148 / 4.146 \\ \hline
		TRLRF & 0.973 / 27.45 & 0.956 / 26.95 & 0.934 / 26.34 & 1.144 / 32.24 & 0.977 / 27.55 & 0.969 / 27.33 & 0.997 / 28.1 & 1.447 / 40.71 \\ \hline
		TRMF & 0.382 / 10.76 & 0.419 / 11.78 & 0.265 / 7.426 & 0.217 / 6.05 & 0.565 / 15.91 & 0.56 / 15.75 & 0.338 / 9.492 & 0.291 / 8.139 \\ \hline
		BTMF & 0.226 / 5.828 & 0.218 / 5.604 & 0.221 / 5.615 & 0.218 / 5.591 & 0.308 / 7.79 & 0.303 / 7.681 & 0.304 / 7.794 & 0.303 / 7.764 \\ \hline
		Reg$L_1$ & 0.521 / 14.66 & 0.489 / 13.78 & 0.192 / 5.429 & 0.132 / 3.718 & 0.699 / 19.69 & 0.498 / 14.01 & 0.242 / 6.825 & 0.155 / 4.387 \\ \hline
		BRTF & 0.243 / 6.829 & 0.232 / 6.522 & {0.138 / 3.888} & {0.131 / 3.706} & 0.218 / 6.133 & 0.2 / 5.609 & {0.152 / 4.285} & {0.145 / 4.121} \\ \hline	
		
	\end{tabular}
	\caption{MRE/RMSE scores for imputation of outlier corrupted data (DT), outlier percentage $(o)$ is 5\% and 10\%. }
	
	\label{table:table2}
\end{table*}

\begin{figure}[ht!]
    \centering
    \begin{subfigure}{.4\linewidth}
        {\label{sen1}\includegraphics[clip, trim=2.8cm 0cm 4cm 1.5cm,width=0.9\linewidth]{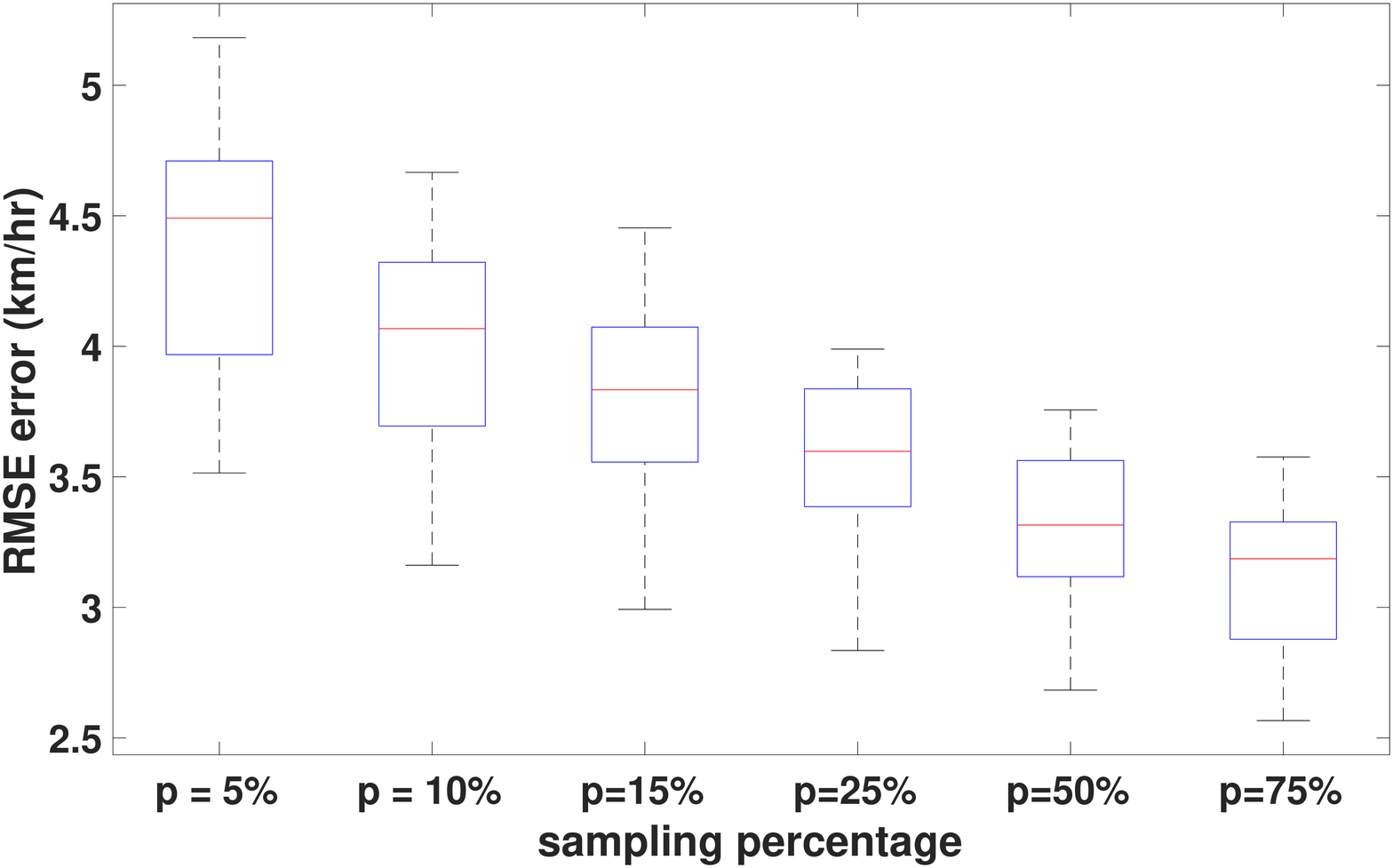}}
        \caption{VBFSI for different different sampling $p=$[0.05, 0.10, 0.25,0.5, 0.75 }
    \end{subfigure}
    \hskip1em
    \begin{subfigure}{.4\linewidth}
        {\label{sen2}\includegraphics[clip, trim=2.5cm 0cm 4cm 1cm,width=0.9\linewidth]{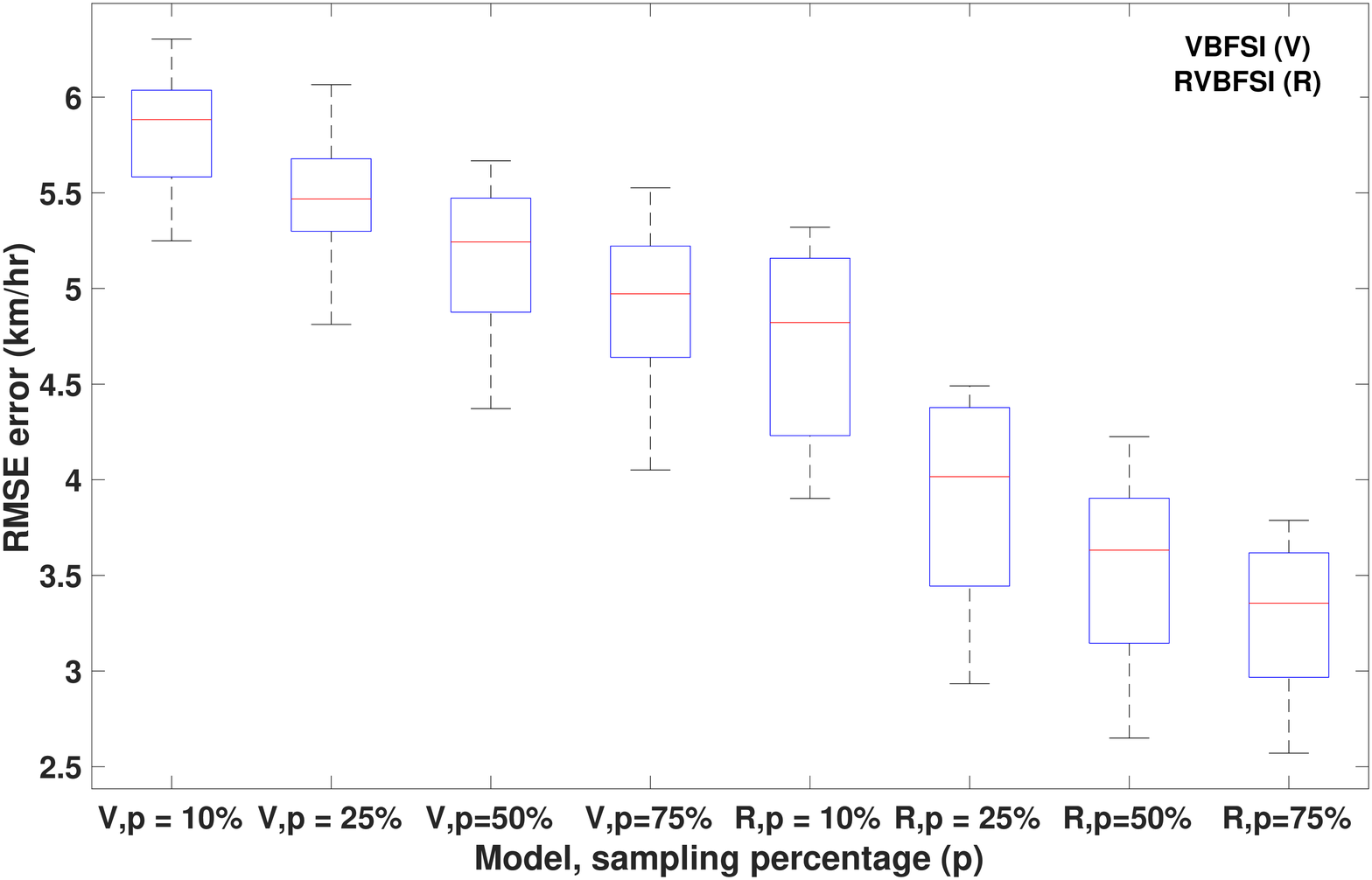}}
        \caption{Comparison of VBFSI and RVBFSI for different $p$ and outlier percentage $o=10\%$}
    \end{subfigure}
    \begin{subfigure}{.4\linewidth}
       {\label{sen3}\includegraphics[clip, trim=2.5cm 0cm 4cm 1cm,width=0.9\linewidth]{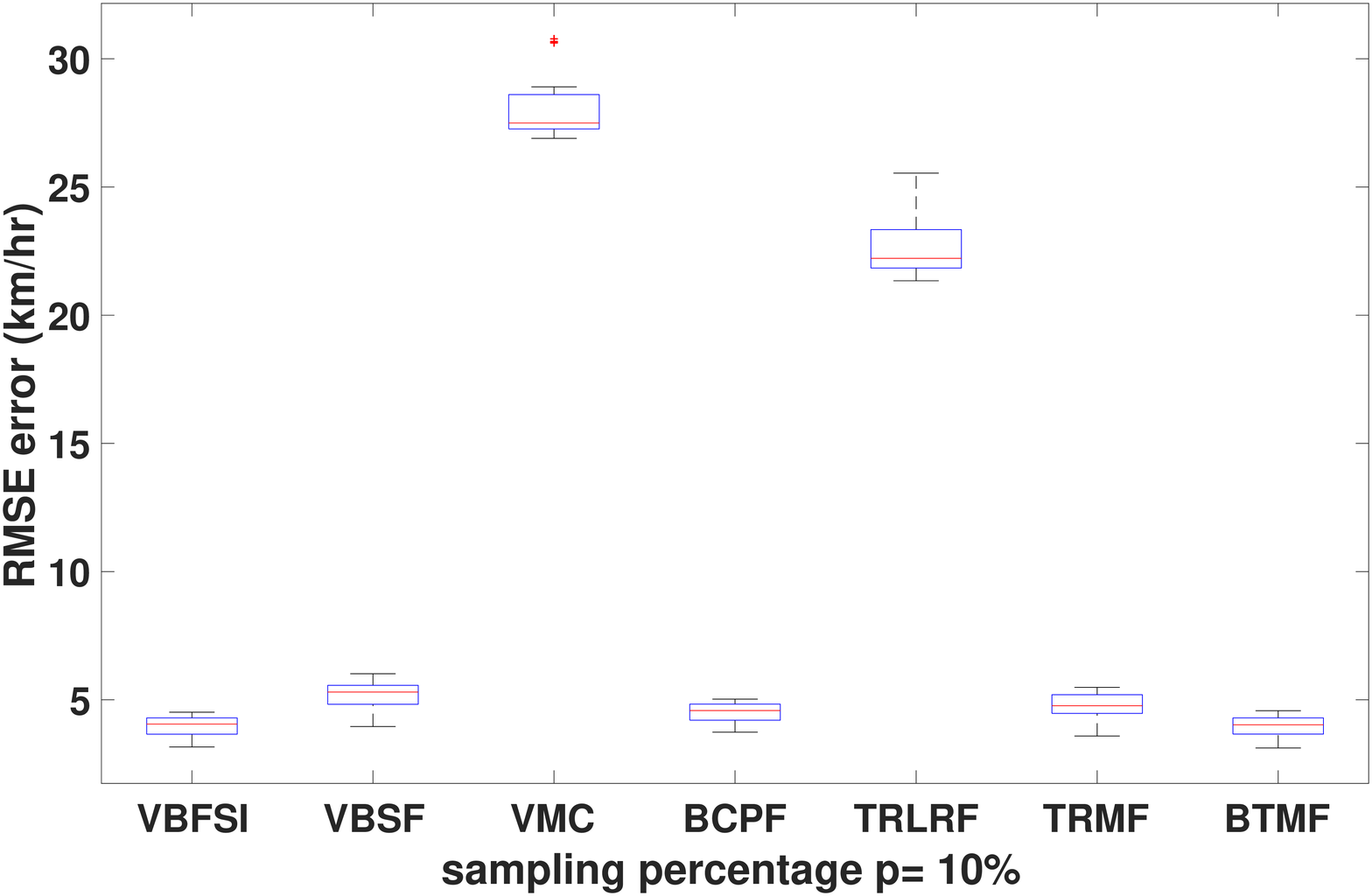}}
        \caption{Imputation performance for  $p=10\%$ }
    \end{subfigure}
    \hskip1em
    \begin{subfigure}{.4\linewidth}
        {\label{sen4}\includegraphics[clip, trim=2.5cm 0.1cm 4cm 1cm,width=0.9\linewidth]{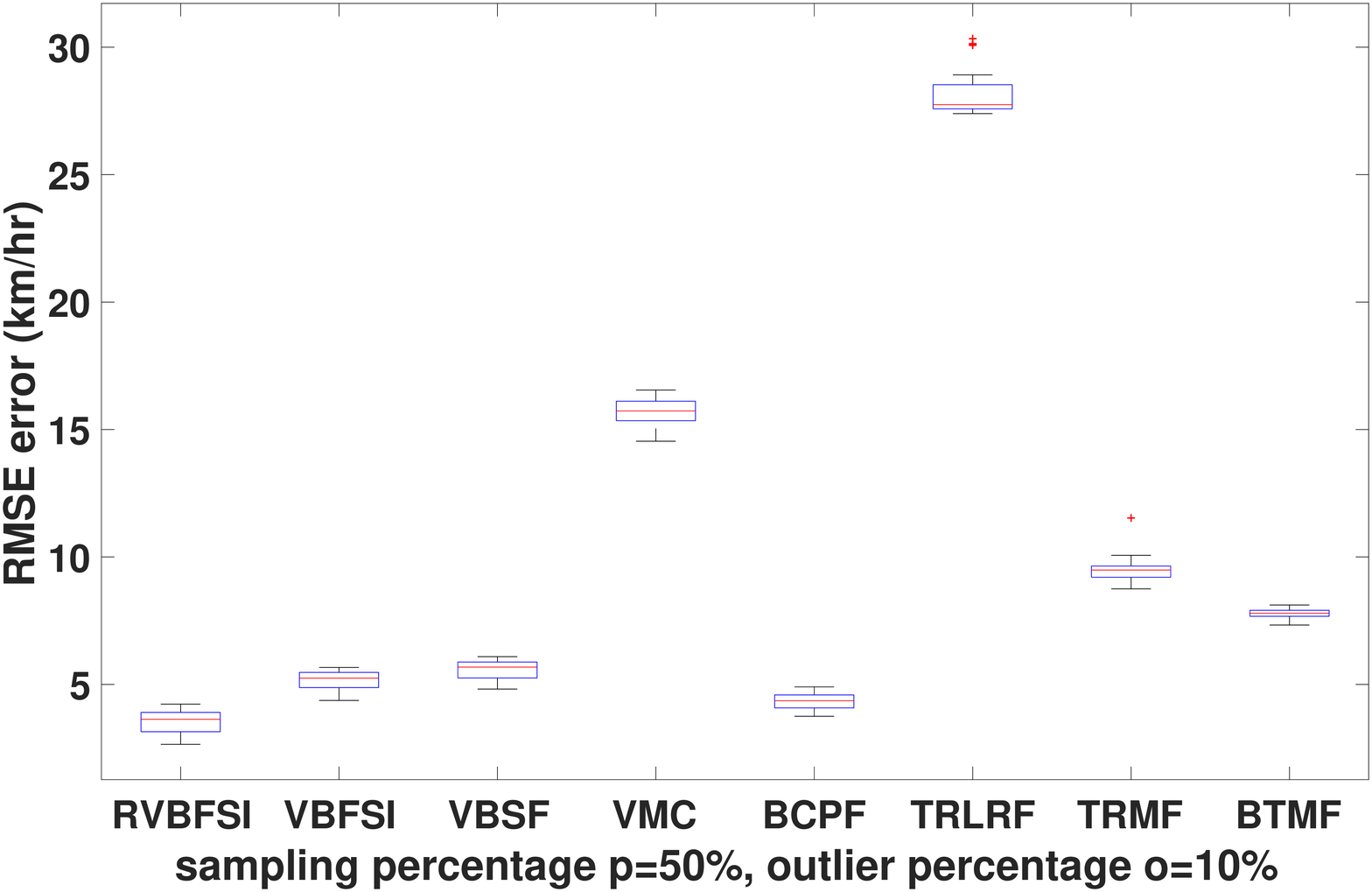}}
        \caption{Imputation performance for $p$=50\% and $o$=10\% }
    \end{subfigure}
   
    \caption{Sensitivity Analysis for Data:DT }
    \label{sen}
\end{figure}

\subsection{Performance Comparison}
The performance comparison of VBFSI with the current state of the art methods is shown in Table  \ref{table:table1}. The performance of RVBFSI for the imputation task for outlier corrupted data is shown in Table  \ref{table:table2}. \\
\textbf{RQ1: Comparison Analysis:}\\
\textbf{Comparison with matrix completion methods}:VBFSI outperforms VBSF for all the datasets. The performance of VBSF is comparable to VBFSI for higher sampling. In contrast, for lower sampling, the performance of VBSF degrades. VMC experience a similar trend, where the performance is comparable for higher sampling and degrades for low sampling. VBFSI outperforms VMC for almost all the cases for traffic data (DT, PT, GT). However, for the air quality data (CA), the performance of VMC is better than VBFSI for a higher sampling percentage. VMC can capture the nonlinearity in the data for a high sampling percentage. 
\begin{figure*}[ht]
	\centering
	\includegraphics[clip, trim=2.5cm 1cm 2.5cm 1.5cm,width=0.85\linewidth]{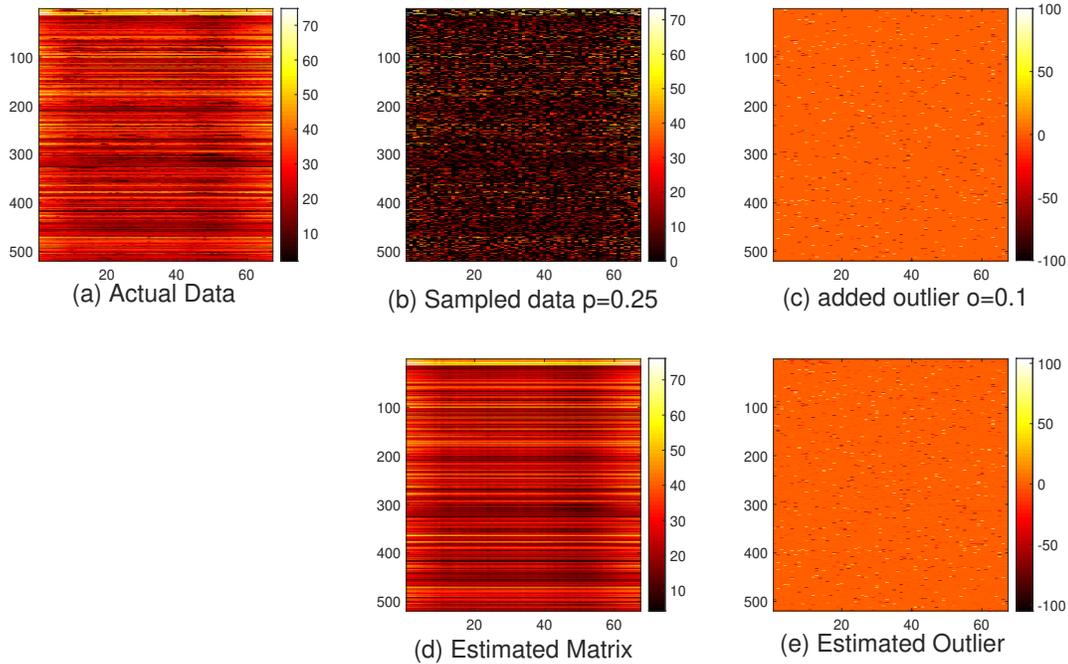}
	\centering
	\caption{\small Robust Matrix completion using RVBFSI for $p$=25\% and $o$=10\% (a) Actual Traffic data (DT) matrix $\X$, (b) $\X$ is sampled with 25 \% of the entries, (c) 10\% of the sampled locations will be corrupted with depicted outlier magnitude and location , (d) Sampled Matrix with outliers are estimated using RVBFSI (e)  Sparse outlier matrix estimated using RVBFSI }
	\label{fig:robust}
\end{figure*}

\textbf{Comparison with Matrix completion frameworks exploiting previous days information}:
For low sampling percentage, VBFSI performance is comparable to BTMF in most of the cases. However, for higher sampling percentage, VBFSI outperforms BTMF. VBFSI outperforms TRMF in all the scenarios. One of the disadvantage of BTMF and TRMF is that rank is not tuned automatically. Moreover BTMF uses gibbs sampling to impute the tensor along with the temporal regularization. Gibbs sampling is relatively slower than the Variational Bayesian approach for parameter estimation\cite{salimans2015markov}.\\
\textbf{Comparison with tensor completion methods}:
VBFSI outperforms TRLRF for all the scenarios. VBFSI performance is comparable to BCPF in most of the cases. \\

\textbf{ RQ2: Effect of $\eta$ on the performance of VBFSI}\\
When $\eta=0$, VBFSI reduces to VBSF. For higher sampling  the performance of VBFSI is comparable to VBSF. However, when the sampling is low, then the performance of VBSF degrades. Incorporating even the noisy prior subspace information in the architecture can reduce the sampling complexity of the matrix by logarithmic factor \cite{prove1}. Therefore, for low sampling VBFSI performs better than VBSF since we have incorporate the prior information in the architecture using $\eta$. 
For lower sampling, value of $\eta$  is high and it decreases exponential with the increase in sampling as shown in Fig. \ref{fig:hyper}.\\


\textbf{RQ3: Performance analysis in the case of Outlier}\\
To compare the performance of VBFSI and RVBFSI in the case of outliers, we artificially add the outliers in the Data: DT. We randomly add 5\% and 10\% of the outliers in the total sampled data, i.e., the number of outliers is $o\times p$ fraction of the overall data. The entries corrupted with outliers are uniformly distributed between  $[-\sigma, \sigma]$, where $\sigma$  is set as 100 in our experimentation.  The imputation performance of VBFSI degrades in the presence of outliers (Fig. \ref{sen})(b). However, RVBFSI can improve the performance of imputation, as shown in Fig. \ref{sen}(b). Moreover, the locations and magnitude of the outliers can be recovered effectively by RVBFSI, as shown in Fig. \ref{fig:robust}(e). Performance comparison of RVBFSI and VBFSI with other imputation methods are shown in Table \ref{table:table2}. The performance of VBFSI and BCPF is comparable for all the sampling. In comparison, the performance of VBFSI is similar to VBSF for high sampling. While RVBFSI outperforms all other imputation methods, including robust imputation methods RVBSF, Reg$\text{L}_1$ and BRTF significantly.

\section{Conclusion}
In this paper, we propose Variational Bayesian Filtering with Subspace Information for the imputation of Spatiotemporal matrices that works even for extreme matrix completion. VBFSI simultaneously models the low rank, temporal evolution, and periodic evolution in one framework to capture the structure in the spatiotemporal data. We incorporate the prior subspace in our model to capture the periodic evolution in the data. We also propose a Robust VBFSI for the imputation of missing data in the presence of outliers. It is observed that RVBFSI performs significantly better than the other imputation methods in the presence of the outliers. 

	\bibliographystyle{IEEEtran}

\bibliography{vbfsi.bib}
\end{document}


\title{Appendix}
\date{}

\maketitle
The update equations for  $\V$ are
\begin{align}
\mub^\V &= \Xib^\V\begin{bmatrix} \hat{\beta}\sum_{i|(i,1)\in\Omega}Z_{i,1}\mub^\A_i  + \Lam_1^{-1}\mub_1\\
\hat{\beta}\sum_{i|(i,2)\in\Omega}Z_{i,2}\mub^\A_i \\
\vdots\\
\hat{\beta}\sum_{i|(i,t)\in\Omega}Z_{i,t}\mub^\A_i
\end{bmatrix} &= \Pb^{-1} \v \label{mub}
\end{align}
\begin{align}
\Pb&=\left[\Xib^{\V}\right]^{-1} &=  \hat{\beta}\Diag{\Xib^\A_{(1)}, \ldots, \Xib^\A_{(t)}}  +
& \begin{bmatrix} \Lam_1^{-1} & -\hat{\J} & \ldots &0\\
-\hat{\J} & \I_r + \Sig^\J & -\hat{\J}  & \ldots \\
\vdots & \vdots &&\vdots \\
\ldots & 0 & -\hat{\J} &  \I_r 
\end{bmatrix} \label{xib}
\end{align}
$\Xib^\B$ is a dense ($tr \times tr$) covariance matrix, and from (22), deriving it would be computationally very expensive $\mathcal{O}^{(t r)^3}$.(Luttinen 2013) has proposed a smoothing algorithm that saves on this computational cost. Since only the diagonal and super-diagonal blocks of $\Xib^\B$ are needed in the iterations, i.e. correlation structures of each row of $\V$ and covariance among neighboring rows, the entire dense matrix $\Xib^\B$ need not be computed. 

The approach followed for inverting $\Pb = (\Xib^\B)^{-1}$ is realizing that it is a symmetric positive-definite matrix. Hence, $\Pb$ may be decomposed via block-LDL decomposition $\Pb = \textrm{LDL}^T$ . Multiplication with inverse of matrix Ψ is equivalent to multiplication with $\textrm{L}^{- T}\textrm{D}^{- 1}\textrm{L}^{- 1}$. To minimize the number of matrix inversions needed, this multiplication is done in two steps:\\
Forward Pass of smoothing encompasses left multiplication with $\textrm{D}^{- 1}\textrm{L}^{- 1}$.Note in the above algorithm, only diagonal and super-diagonal block matrices of $\Pb$ need be computed (Algorithm \ref{Algo 1}).\\
Backward Pass of smoothing encompasses left-multiplication with $\textrm{L}^{- T}$. This operation is performed on the outputs of Forward Pass (Algorithm \ref{Algo 2}). The algorithm is in-place, i.e.
new variables need not be created in Backward Pass.

\begin{algorithm}
	\caption{FORWARD PASS}

	\textbf{Input}:(\Pb_{k,k}|_{k=1}^{t},\Pb_{k,k+1}|_{k=0}^{t-1},\v_{k}|_{k=1}^t )
	
	\begin{algorithmic}[1] 
	 \STATE	$\hat{\Xib}_{0,0}=\Pb_{0,0}^{-1}$ \\
	 \STATE	$\mub_0=\hat{\Xib}_{0,0} \v_0$
		\FOR{$k=1$ to $t-1$}
		
		\STATE $\hat{\Xib}_{k,k+1}= \hat{\Xib}_{k,k} \Pb_{k,k+1}$
		\STATE $\hat{\Xib}_{k+1,k+1}=( \Pb_{k+1,k+1}- \hat{\Xib}_{k,k+1}^T \Pb_{k,k+1} )^{-1}$
		\STATE $\hat{\mub}_{k+1}=\hat{\Xib}_{k+1,k+1}(\v_{k+1}- \hat{\Xib}_{k,k+1}^T \hat{\mub}_{k} )$
		
		\ENDFOR
		\STATE 
		\textbf{Output}: $\hat{\mub}_{k}|_{k=1}^t, \,\, \hat{\Xib}_{k,k+1}|_{k=0}^{t-1},\,\, \hat{\Xib}_{k,k}|_{k=1}^{t} $
	\end{algorithmic}
	\label{Algo 1}
\end{algorithm}

\begin{algorithm}
	\caption{BACKWARD PASS}

\textbf{Input}:($\hat{\mub}_{k}|_{k=1}^t, \,\, \hat{\Xib}_{k,k+1}|_{k=0}^{t-1},\,\, \hat{\Xib}_{k,k}|_{k=1}^{t}$ )

\begin{algorithmic}[1] 
	\STATE	 {\Xib}_{t,t}=\hat{\Xib}_{t,t}, \,\,
	\STATE	$\mub_0=\hat{\mub}_t$
	\FOR{$k=1$ to $t-1$}
	
	\STATE ${\Xib}_{k,k+1}= -\hat{\Xib}_{k,k+1} {\Xib}_{k+1,k+1}$
	\STATE ${\Xib}_{k,k}=  \hat{\Xib}_{k,k}-  \hat{\Xib}_{k,k+1} {\Xib}_{k,k+1}^T$
	\STATE ${\mub}_{k}=\hat{\mub}_{k} - \hat{\Xib}_{k,k+1} \mub_{k+1} $
	
	\ENDFOR
	\STATE 
	\textbf{Output}: ${\mub}_{k}|_{k=1}^t, \,\, {\Xib}_{k,k+1}|_{k=0}^{t-1},\,\, {\Xib}_{k,k}|_{k=1}^{t} $
\end{algorithmic}
\label{Algo 2}
\end{algorithm}

Overall updates for $\V$ using the FORWARD PASS and BACKWARD PASS is as follows:
\begin{align}
(\hat{\mub}_{k}, \,\, \hat{\Xib}_{k,k+1},\,\, \hat{\Xib}_{k,k}) \leftarrow \text{FORWARD PASS}( \Pb_{k,k},\Pb_{k,k+1},\v)
\end{align}

\begin{align}
({\mub}_{k}, \,\, {\Xib}_{k,k+1},\,\, {\Xib}_{k,k}) \leftarrow \text{BACKWARD PASS}( \hat{\mub}_{k}, \,\, \hat{\Xib}_{k,k+1},\,\, \hat{\Xib}_{k,k})
\end{align}